\documentclass{emulateapj}

\newcommand{\teff}{$T_{\rm eff}$ }
\newcommand{\tsin}{$T_{\rm eff}$}
\newcommand{\tef}{T_\mathrm{eff}}
\newcommand{\logg}{$\log g$}
\newcommand{\feh}{$\mathrm{[Fe/H]}$}

\slugcomment{Submitted to ApJ}

\shorttitle{A signature of the rocky core in 16 CygBb}
\shortauthors{Maia, Mel\'endez and Ramz\'irez}

\begin{document}

\title{High precision abundances in the 16 Cyg binary system: a signature of
the rocky core in the giant planet}

\author{
Marcelo Tucci Maia\altaffilmark{1},
Jorge Mel\'{e}ndez\altaffilmark{1} and
Iv\'an Ram{\'{\i}}rez\altaffilmark{2},
}
\email{marcelotuccimaia@usp.br}
\altaffiltext{1}{Departamento de Astronomia do IAG/USP, Universidade de S\~ao Paulo, Rua do Mat\~ao 1226, 
Cidade Universit\'aria, 05508-900 S\~ao Paulo, SP, Brazil}
\altaffiltext{2}{McDonald Observatory and Department of Astronomy, University of Texas at Austin, TX, USA}
\altaffiltext{*}{Based on observations obtained at the
Canada-France-Hawaii Telescope (CFHT) at the 3.6m telescope at Mauna Kea.}

%\altaffiltext{1}{Visiting Astronomer, Magellan Telescopes, Las Campanas Observatory, Chile,
%and W.M. Keck Observatory, Manua Kea, Hawaii}


\begin{abstract}
We study the stars of the binary system 16~Cygni to determine with high precision their chemical composition. 
Knowing that the component B has a detected planet of at least 1.5~Jupiter masses, we investigate if there are 
chemical peculiarities that could be attributed to planet formation around this star.
We perform a differential abundance analysis using high resolution ({\it R}=81,000) and high 
S/N ($\sim$700) CFHT/ESPaDOnS spectra of the 16~Cygni stars and the Sun; 
the latter was obtained from light reflected of asteroids.  
We determine differential abundances of the binary components relative to the Sun and 
between components A and B as well.
We achieve a precision of $\sigma$ $\lesssim$0.005 dex and a total error $\sim$0.01 dex for 
most elements. The effective temperatures and surface gravities found for 16~Cyg~A and B 
are \teff = 5830$\pm$7 K, \logg = 4.30$\pm$0.02 dex, and \teff = 5751$\pm$6 K, \logg = 4.35$\pm$0.02 dex, 
respectively. The component 16~Cyg~A has a metallicity (\feh) higher by 0.047$\pm$0.005 dex 
than 16~Cyg~B, as well as a microturbulence velocity higher by 0.08 km s$^{-1}$.
All elements show abundance differences between the binary components,
but while the volatile difference is about 0.03 dex, the refractories differ by more
and show a trend with condensation temperature, which could be interpreted as the signature of the
rocky accretion core of the giant planet 16~Cyg~Bb. 
We estimate a mass of about 1.5-6 {\it M}$_\Earth$ for this rocky core, in good
agreement with estimates of Jupiter's core.
\end{abstract}


\keywords{planetary systems ---  stars: abundances --- Sun: abundances}

\section{Introduction}

It is common to assume that stars of multiple stellar systems have the same chemical composition, 
since they originated from the same natal cloud. However, some studies indicate that, 
in binary systems, there may be small differences in the chemical composition of their components
\citep{gra01,la01,des04,des06,ram11}. 
One explanation for these anomalies is planet formation \citep[e.g.,][]{la01,ram11}.

The binary system 16 Cygni is known for having a detected giant planet
orbiting the B component, with a minimum mass of 1.5 {\it M}$_{\rm Jup}$ \citep{coc97} and
a probable true mass of about 2.4 {\it M}$_{\rm Jup}$ \citep{pla13}. Even though the system has been
monitored for small radial velocity variations for over two decades, so far no planets have been detected 
around the primary, which makes this system ideal to study the
formation of giant planets. However, the chemical signatures of planet formation on the host star 
are expected to be very small, of only a few 0.01 dex \citep{mel09,ram09,cham10}, hence
a high precision is needed to detect these effects.

Although earlier analyses of the 16 Cyg system suggested that 16 Cyg A is
about 0.05 dex more metal-rich than 16 Cyg B \cite[e.g.][]{gon98}, the difference
is so small that it could be due to the relatively large abundance uncertainties
of these earlier studies.
In a pioneer precise line-by-line differential study of this binary, \cite{la01} found a difference 
(A - B) of +0.025$\pm$0.009 dex in the iron abundance of both components.
Seeking for potential additional signatures of giant planet formation, \cite{ram11} performed 
a differential abundance determination of 25 elements 
and discovered significant differences among all chemical elements that were analyzed, 
with component A being more metal rich by 0.04$\pm$0.01 dex than B. In contrast, in a study published at about 
the same date, \cite{sch11} found no difference in the chemical composition of these two stars.
The intent of this work is to shed more light into this matter using better quality spectra and discuss 
the possible chemical signature caused by the formation of gas giant planets.

\section{Observations and Data Reduction}

Spectra of 16 Cyg A and B were obtained with the Echelle SpectroPolarimetric Device for 
Observation of Stars (ESPaDOnS) on the 3.6 m Canada-France-Hawaii Telescope (CFHT) at Mauna Kea. 
The observations took place on 2013 June 06 on Queued Service Observing (QSO) mode. 
The observations were taken with the fiber only on the object (Spectroscopy, star o), 
that is the highest resolution ({\it R} = 81,000) on the instrument. Notice
that our resolving power is significantly higher than that used in the previous 
studies of \cite{sch11} and \cite{ram11}, {\it R} = 45,000 and {\it R} = 60,000, respectively.

The exposure times were 3$\times$280 and 3$\times$350 s on 16 Cyg A and B, respectively,
with 16 Cyg B observed immediately after 16 Cyg A.
We achieved a S/N $\sim$700 around 600 nm for each of the binary components. 
The asteroids Vesta and Ceres were also observed with the same spectrograph setup 
to acquire the solar spectrum that served as the reference in our differential analysis. 
A similar S/N ($\sim$700) was achieved for both asteroids. Our S/N ratios are higher
than those obtained by \cite{ram11}, S/N $\sim$ 400, and about the same as that obtained
by \cite{sch11} for 16 Cyg A (S/N = 750).

We used the pipeline reduced spectra provided by CFHT, which passed through the usual 
reduction process including bias subtraction, flat fielding, spectral order extractions 
and wavelength calibration. We performed the continuum normalization of the spectra using IRAF.

\section{Analysis}

We used the line-by-line differential method to obtain stellar parameters and
chemical abundances, as described in \cite{mel12} and \cite{mon13}. 
The 2002 version of the LTE code MOOG \citep{sne73} was used with 
Kurucz ODFNEW model atmospheres \citep{cas04}.

The adopted line list is an updated version of that presented in \cite{mel12},
with several dozen lines added.
The equivalent width (EW) measurements were made  by hand with the task splot in IRAF,  
using Gaussian profile fits. The local continuum was
carefully selected by over-plotting the spectra of both binary components 
and the solar spectrum for each line.

We obtained the abundance of 18 elements: C, O, Na, Mg, Al, Si, S, Ca, Sc, 
Ti, V, Cr, Mn, Fe, Co, Ni, Cu, and Zn. All abundances were differentially 
determined line-by-line using the Sun as standard in a first approach and 
then using 16 Cyg B as reference to obtain the 16 Cyg A - B ratios. 
The differential method minimizes errors 
due to uncertainties in the line transition probabilities 
and shortcomings of model atmospheres, allowing thus
an improved determination of stellar parameters and chemical abundances.
The elements V, Mn, Co, and Cu had their abundances 
corrected for hyperfine structure (HFS). For this calculation 
the blends driver in MOOG was used adopting the HFS data from \cite{mel12}.

The atmospheric parameters for 16 Cyg A and B were obtained by 
differential excitation equilibrium (for \tsin) and differential 
ionization equilibrium (for \logg), using as reference solar abundances
for FeI and FeII lines.
First, we determined absolute abundances for the Sun using the solar atmospheric 
parameters of 5777 K for \teff and 4.44 for \logg, and adopting an
initial microturbulence velocity of $v_t$ = 0.9 km$/s^{\rm -1}$. Then, we 
estimated $v_t$ by the usual method of requiring zero slope in
the absolute abundances of FeI lines versus reduced EW. We obtained a final  
$v_t$ = 0.86 km$/s^{\rm -1}$ for the Sun, and computed  our reference solar abundances
for each line.

The next step was the determination of stellar parameters for the 16 Cygni stars. 
Initially, we used model atmospheres with the 
parameters published in \cite{ram11}: 
\teff = 5813 K, \logg = 4.28 and [Fe/H] = 0.10 for 16 Cyg A, 
and \teff = 5749 K, \logg = 4.33 and [Fe/H] = 0.06 for 16 Cyg B.  
Then, we iteratively changed the stellar parameters of 16 Cyg A and
B until we achieved the differential excitation and ionization equilibrium,
and also no trend in the differential FeI abundances with reduced EW
(to obtain $v_t$), changing the metallicity of the models at each 
iteration until reaching convergence. 

Our derived stellar parameters using the Sun as a standard are \teff = 5830$\pm$11 K, 
\logg = 4.30$\pm$0.02, $v_t$ = 0.98$\pm$0.02 km$/s^{\rm -1}$ and [Fe/H] = 0.101$\pm$0.008 dex
for 16 Cyg A, and \teff = 5751$\pm$11 K, \logg = 4.35$\pm$0.02, $v_t$ = 0.90$\pm$0.02 km$/s^{\rm -1}$ 
and [Fe/H] = 0.054$\pm$0.008 dex for 16 Cyg B.
These errors take into account the errors in the measurements and 
the degeneracy of stellar parameters. 
A similar procedure was repeated but using 16 Cyg B as the reference star 
instead of the Sun to perform the differential spectroscopic equilibrium (Figure 1), 
and fixing the stellar parameters of the B component
to our results from the differential analysis relative to the Sun. 
The resulting atmospheric parameters for the A component are the same 
as when the Sun is used as a reference,
but with smaller errors for \teff ($\pm$ 7 K) and $v_t$ ($\pm$0.01 km$/s^{\rm -1}$). 
The final $\Delta$(Fe) difference for 16 Cyg A minus 16 Cyg B is 0.047 $\pm$ 0.005 dex, 
confirming that there is indeed a difference in the metallicity between the two stars 
of this binary system. 

\begin{figure}
\centering
\includegraphics[width=\columnwidth]{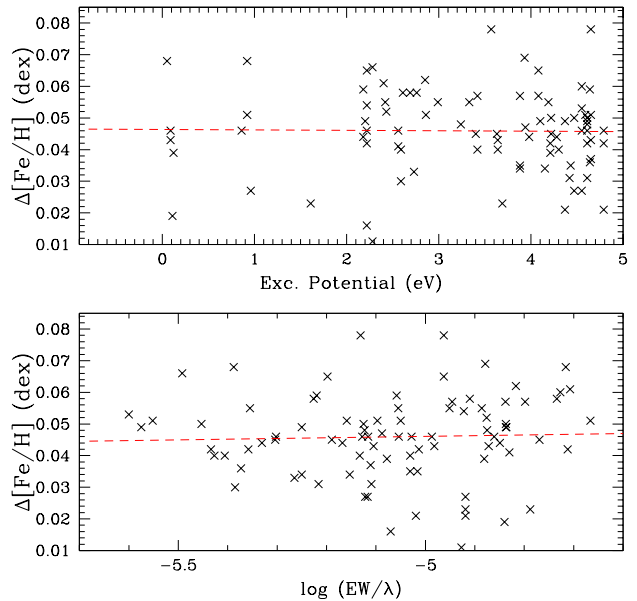}
\caption{Differential FeI abundances (16 Cyg A - 16 Cyg B) as a function of 
excitation potencial (upper panel) and reduced EW (bottom panel).}
\label{fig1}
\end{figure}

Our stellar parameters are in very good agreement with the ones 
determined by \cite{ram11}. We obtain somewhat higher both effective temperatures
and log $g$  by 17 K and 0.02 dex in the case of 16 Cyg A, 
and by 2 K and 0.03 dex for 16 Cyg B. The relative
difference between the components A and B is in even better agreement,
with our results showing a $\Delta$\teff higher by 15 K than
\cite{ram11}, and the difference in the $\Delta$log $g$ is 0.00 dex.
Using the photometric IRFM calibrations of \cite{ram05} for 13 different
optical and infrared colors (Table \ref{colors}) from the Jhonson,  Cousins, Vilnius, Geneva, DDO, Tycho and 2MASS
systems \citep{tay86,mer97,hog00,cut03}, with the corresponding [Fe/H] for each binary component,
we determined average, median and trimean\footnote{The trimean is a robust estimate of central tendency.
We define trimean = (Q1 + 2$\times$median + Q3)/4), where Q1 and Q3 are the first and third quartile.}
effective temperatures (Table \ref{colors}) for the binary pair,
resulting in a  temperature difference of $\Delta T_{\rm eff}^{\rm phot}$ (A-B) = 58$\pm$10, 78$\pm$10, 73$\pm$10 K,
for the difference of average, median and trimean temperatures. The two robust indicators, median and trimean,
are in excellent agreement with our spectroscopic $\Delta T_{\rm eff}^{\rm spec}$ (A-B) = 79$\pm$7 K, and also
in agreement with the results from \cite{ram11}, who found $\Delta T_{\rm eff}^{\rm spec}$ (A-B) = 64$\pm$25 K.
Compared to \cite{sch11}, our $\Delta$\teff and $\Delta$log $g$ are higher
by +36 K and +0.03 dex, respectively. Notice that according to the 
trigonometric log $g$ \citep{ram11}, $\Delta$log $g$ should be 0.05 dex
between the components, that is the value found in our work and by
\cite{ram11}, but \cite{sch11} found a lower $\Delta$log g = 0.02 dex,
although our results are in agreement with \cite{sch11} within their error bars.

Once the stellar parameters of the 16 Cygni stars were set using iron lines,
we computed abundances for all remaining elements.
In Table \ref{parameters1} we present the final differential abundances 
of 16 Cyg A relative to 16 Cyg B, and their respective errors, 
while in Table \ref{parameters2} we present the abundances and errors for 
16 Cyg A and B using the Sun as standard.  
We present both the observational errors and systematic errors due
to uncertainties in the stellar parameters, as well as the total
error obtained by adding quadratically both errors.

\section{Results and Discussion}

The differential abundances of the 16 Cyg pair relative to the Sun
are shown in Figure 2. Both 16 Cyg A and B show abundances that have a 
clear trend with condensation temperature,
as already shown by \cite{ram11} and \cite{sch11}. There is a reasonable
agreement with the mean trend of 11 solar twins relative to the Sun by \cite{mel09},
shown by solid lines  in Figure 2, after a vertical shift is applied to 
match the refractory elements. Interestingly, the same qualitative 
pattern as in \cite{ram11} is found for individual volatile elements in both components, 
with O somewhat higher than C, and Zn somewhat higher than S. Thus, the variations
among the volatile elements are likely real.

\begin{figure}
\centering
\includegraphics[width=\columnwidth]{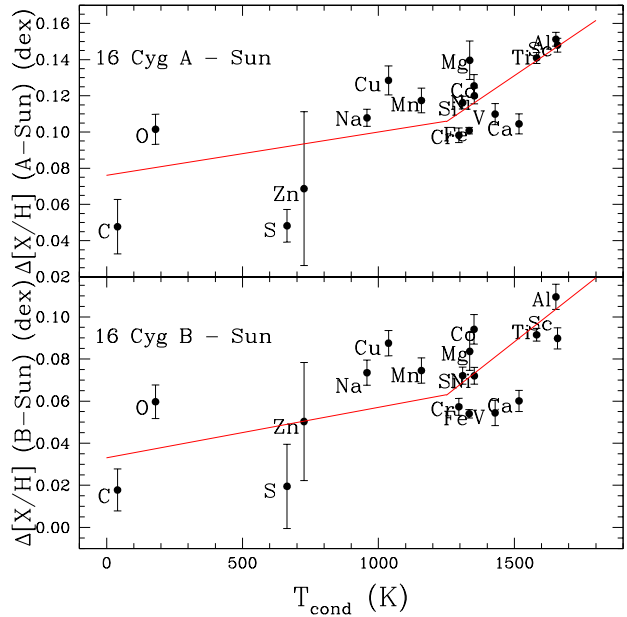}
\caption{Differential abundances of 16 Cyg A - Sun (top panel) and
16 Cyg B - Sun (bottom panel) vs. condensation temperature.
The solid lines are the mean trend determined by \cite{mel09} for 
11 solar twins compared to the Sun, after vertical shifts are applied
to match the highly refractory elements in 16 Cyg - Sun.}
\label{fig2}
\end{figure}

From Figure 2, it is already noticeable that there are abundance differences
between the two 16 Cygni components, with 16 Cyg A being more metal-rich. 
The differential abundances of 16 Cyg A relative
to 16 Cyg B, plotted in Figure 3, shows this more clearly.
As already found by \cite{ram11}, all elements seem enhanced in 16 Cyg A,
but now this is more evident due to our higher precision.
This is contrary to the results obtained by \cite{sch11}, who found 
no chemical difference in the binary pair.
The differential analysis of \cite{tak05} also showed both components 
to have the same iron abundance, but the S/N of his spectra 
(S/N$\sim$100) is too low for a precise analysis.

\cite{ram11} found a roughly constant difference of about 0.04 dex in 
the differential abundances (A - B) of volatiles and refractories.
However, while in our study the volatile elements show a difference of about 0.03 dex,
the refractories show larger differences and a trend with condensation
temperature (Figure 3). A similar trend has been
reported in a short note added in proof by \cite{la01}, where based
on the analysis of 13 elements, a correlation with condensation temperature
is found, with a slope of 1.4 $\pm$ 0.5$\times$10$^{-5}$ dex K$^{-1}$, however, no
further details are given. Interestingly, the same slope of A minus B (1.4$\pm2.8\times10^{-5}$) 
is found by \cite{sch11}.
In this work we obtain a slope for 
the refractories of 1.88  $\pm$ 0.79$\times$10$^{-5}$ dex K$^{-1}$, in reasonable
agreement with the results by \cite{la01} and \cite{sch11}.
Notice that the abundance difference that we find here for 16 Cyg A - B,
is very distinct from the mean trend for the 11 solar twins of \cite{mel09}, shown
by a dot-dashed line in Figure 3 after a shift has been applied to fit highly refractory elements.

\begin{figure}
\centering
\includegraphics[width=\columnwidth]{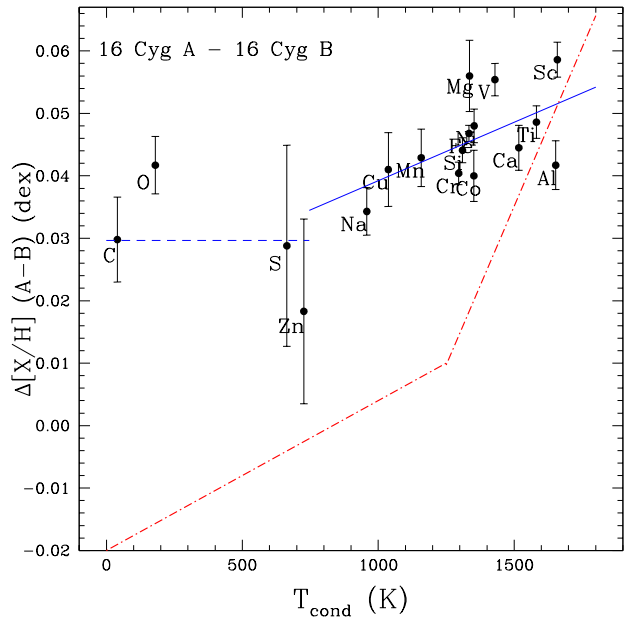}
\caption{Differential abundances of 16 Cyg A - 16 Cyg B versus
condensation temperature. The dashed line is the average of
the volatiles and the solid line the trend of the refractories.
The dot-dashed line is the mean trend obtained by \cite{mel09} for 
11 solar twins compared to the Sun, after a vertical shift is applied
to match the highly refractory elements in A - B.}
\label{fig3}
\end{figure}

The overall deficiency in the abundances of 16 Cyg B (compared to 16 Cyg A),
could be attributed to the formation of its giant planet,
as the metals missing in 16 Cyg B could have been taken from the proto-planet disk
to form its gaseous giant. More interestingly, for the refractories
we seem to detect a trend with condensation temperature, that may
represent the detection, for the first time, of a signature of the rocky accretion core
of a giant planet. In the core accretion model \cite[e.g.,][]{pap06}, first a rocky core forms
through coagulation of planetesimals until it is massive enough for
accretion of a gas envelope, to form a giant planet.
The higher deficiency of refractories in 16 Cyg B, means that the
giant planet 16 CygBb may have an excess of refractories, which could be
due to its rocky accretion core. Another important hint is that the
break in condensation temperature between volatiles and refractories
seem to occur around T$_{\rm cond}$ $\sim$ 500-700 K. This means that most
likely the rocky core was not formed in the inner disk regions 
(equivalent to the Earth-Sun distance) but it was formed at a larger distance, 
where giant planets are more likely to form.

Following \cite{cham10}, we estimate the mass of the rocky core by adding  
a mixture of the composition of the Earth and CM chondrites to the convective zone 
of 16 Cyg B. Assuming a similar convection zone as in the Sun (0.023 M$_\odot$), 
we can reproduce the trend of the refractories (Figure \ref{fig3}) by adding about 1.5 M$_\Earth$ of 
rocky material with the above composition. Notice that this is the minimum mass of 
the rocky core because we do not know the exact size of the convection zone at the
time the giant planet was formed \citep[see discussion in][]{ram11}.
Assuming a convection zone of about 0.1 {\it M}$_\odot$ at the time of the formation of
the giant planet's core, the mass of the rocky core would be higher (6 {\it M}$_\Earth$).
Our estimates (1.5 - 6 {\it M}$_\Earth$) are consistent with Jupiter's core mass of $\sim$ 5$\pm$5 {\it M}$_\Earth$
\citep{gui05}.

\section{Conclusions}

We found significant differences in the chemical abundances of 16 Cyg A relative to 16 Cyg B, 
for all analysed elements. The abundance differences range from 0.03 dex for the volatiles 
up to 0.06 dex for the refractories.

The 16 Cygni system is so far a unique case where high-precision abundance analyses
show a sharp distinction in the chemical composition of the binary components.
A recent study of the binary pair HAT-P-1, 
where the secondary hosts a giant planet of 0.53 M$_{\rm Jup}$ \citep{bak07}
but no planets have been detected so far around the primary,
reveals no abundance contrast \citep{liu14}. One explanation for the lack of abundance differences 
between the binary components of HAT-P-1 could be that the mass of the planet is
much smaller than the planet in the 16 Cygni system, which has about 2.4 {\it M}$_{\rm Jup}$ \citep{pla13}.
Scaling by the mass of the planets, the observed difference of about 0.04 dex in the chemical
abundances of the 16 Cygni pair, would imply in a dissimilarity of only 0.009 dex for the HAT-P-1 binary, 
which would be challenging to detect. Another recent study of a binary pair, HD 20781/HD 20782, 
where HD 20782 has a Jupiter-mass planet and HD 20781 hosts two Neptune-mass planets, 
show zero abundance differences (0.04 $\pm$ 0.07 dex) within the error bars \citep{mac14}.

In any case, our findings could be interpreted as due to the formation of the giant planet 
around 16 Cyg B. Within that scenario, we have detected, for the first time, the signature 
of the rocky accretion core of the giant planet 16 Cyg Bb, with a mass of $\sim$ 1.5-6 {/it M}$_\Earth$.
Our study opens new windows on the study of the planet - star connection.

\acknowledgments

MTM thanks support by CAPES.
JM thanks support by FAPESP (2012/24392-2) and CNPq (Bolsa de produtividade).

\begin{table}
\begin{minipage}[t]{\textwidth}
\caption{Colors of 16 Cyg A and B on Different Photometric Systems, and their Corresponding 
Mean Effective Temperatures using the Calibrations of \cite{ram05}.}
\label{colors}
\centering
\begin{tabular}{lrr} 
\hline\hline                
 {Color}& 16Cyg A & 16Cyg B \\
\hline    

{\it (B-V)}  & 0.644 & 0.663\\
{\it (b-y)}  & 0.410 & 0.416\\
{\it (Y-V)} & 0.569 & 0.575\\
{\it (V-S)} & 0.557 & 0.569\\
{\it ($B_{\rm 2}$-$V_{\rm 1}$)} & 0.398 & 0.402\\
{\it ($B_{\rm 2}$-G)} & 0.109 & 0.117\\
{\it (V-R)$_{\rm C}$} & 0.357 & 0.363\\
{\it (V-I)$_{\rm C}$} & 0.698 & 0.706\\
{\it (R-I)$_{\rm C}$} & 0.341 & 0.343\\
{\it C(}42-45{\it)} & 0.648 & 0.669\\
{\it C(}42-48{\it)} & 1.671 & 1.698\\
{\it ($B_{\rm T}$-$V_{\rm T}$)} & 0.722 & 0.732\\
{\it ($V$-$K_{\rm 2}$)} & 1.533 & 1.577\\
\hline    
$T_{\rm eff}^{\rm average}$ (K) & 5726 & 5668\\
$T_{\rm eff}^{\rm median}$  (K) & 5737 & 5659\\
$T_{\rm eff}^{\rm trimean}$ (K) & 5734 & 5661\\
$\sigma$ (K) & 29 & 23 \\
s.e. (K) & 8 & 6 \\

\hline                                 
\end{tabular}
\end{minipage}
\end{table}

\begin{table}
%\begin{minipage}[t]{\textwidth}
\caption{Differential Abundances of 16 Cyg A - 16 Cyg B, and their Errors}
\label{parameters1}
\centering
\begin{tabular}{llrrrrlll} 
\hline\hline                
 {Element}& LTE   & $\Delta \tef$ & $\Delta$log $g$ & $\Delta v_t$ & $\Delta$[Fe/H] & param\tablenotemark{a} & obs\tablenotemark{b} & total\tablenotemark{c} \\
\hline    
{}       &       & +7K           &  +0.02 dex      & +0.01 km s$^{-1}$  & +0.01 dex   &  &  &  \\
{}       & (dex) & (dex)         & (dex)           & (dex)       & (dex)        & (dex) & (dex) & (dex) \\
\hline    
C  & 0.030 &-0.004 & 0.003 & 0.000  & 0.000 & 0.005 & 0.007 & 0.008\\
O  & 0.042 &-0.006 & 0.002 & -0.001 & 0.002 & 0.006 & 0.005 & 0.008\\
Na & 0.034 &0.003  & -0.001& 0.000  & 0.000 & 0.003 & 0.004 & 0.005\\
Mg & 0.056 &0.004  & -0.001& -0.001 & 0.000 & 0.004 & 0.006 & 0.007\\
Al & 0.042 &0.003  & -0.001& 0.000  & 0.000 & 0.003 & 0.004 & 0.005\\
Si & 0.044 &0.001  & 0.001 & -0.001 & 0.001 & 0.002 & 0.002 & 0.003\\
S  & 0.029 &-0.004 & 0.003 & 0.000  & 0.001 & 0.005 & 0.016 & 0.017\\
Ca & 0.045 &0.004  & -0.001& -0.002 & 0.000 & 0.004 & 0.004 & 0.006\\
Sc & 0.059 &0.005  & 0.000 & -0.001 &-0.001 & 0.005 & 0.003 & 0.006\\
Ti & 0.049 &0.006  & 0.000 & -0.002 & 0.000 & 0.006 & 0.003 & 0.006\\
V  & 0.055 &0.006  & 0.001 & 0.000  & 0.000 & 0.006 & 0.003 & 0.007\\
Cr & 0.040 &0.004  & -0.001& -0.002 & 0.000 & 0.005 & 0.002 & 0.005\\
Mn & 0.043 &0.005  & -0.001& -0.003 & 0.000 & 0.005 & 0.005 & 0.007\\
Fe & 0.047 &0.004  & -0.001& -0.002 & 0.000 & 0.005 & 0.001 & 0.005\\
Co & 0.040 &0.004  & 0.001 & 0.000  & 0.000 & 0.004 & 0.004 & 0.006\\
Ni & 0.048 &0.003  & 0.000 & -0.002 & 0.001 & 0.004 & 0.003 & 0.005\\
Cu & 0.041 &0.003  & 0.001 & -0.002 & 0.001 & 0.004 & 0.006 & 0.007\\
Zn & 0.018 &0.000  & 0.001 & -0.003 & 0.002 & 0.004 & 0.015 & 0.015\\

\hline                                 
\end{tabular}
    \tablenotetext{1}{Errors due to stellar parameters}
    \tablenotetext{2}{Observational errors}
    \tablenotetext{3}{Quadric sum of the observational and stellar parameters uncertainties}
%\end{minipage}
\end{table}

\begin{table}
\begin{minipage}[t]{\textwidth}
\caption{Differential Abundances of 16 Cyg A and B using the Sun as a Standard, and their Errors}
\label{parameters2}
\centering
\begin{tabular}{lllrrrrlll} 
\hline\hline                
 {Element}& 16Cyg A & 16Cyg B   & $\Delta \tef$ & $\Delta$log $g$ & $\Delta v_t$ & $\Delta$[Fe/H] & param\tablenotemark{a} & obs\tablenotemark{b} & total\tablenotemark{c} \\
\hline    
{}       &       &       & +11K           &  +0.02 dex      & +0.02 km s$^{-1}$  & +0.01 dex   &  &  &  \\
{}       & (dex) & (dex) & (dex)         & (dex)           & (dex)       & (dex)        & (dex) & (dex) & (dex) \\
\hline    
C  & 0.048 & 0.018 &-0.005 & 0.003 & 0.000 & 0.000 & 0.006 & 0.015 & 0.016\\
O  & 0.102 & 0.060 &-0.007 & 0.002 & 0.001 & 0.002 & 0.008 & 0.008 & 0.011\\
Na & 0.108 & 0.074 &0.004  & -0.001& 0.001 & 0.000 & 0.004 & 0.005 & 0.006\\
Mg & 0.140 & 0.084 &0.005  & -0.001& 0.003 & 0.000 & 0.006 & 0.011 & 0.012\\
Al & 0.151 & 0.110 &0.003  & -0.001& 0.001 & 0.000 & 0.004 & 0.004 & 0.005\\
Si & 0.116 & 0.072 &0.002  & 0.001 & 0.001 & 0.001 & 0.002 & 0.004 & 0.004\\
S  & 0.048 & 0.020 &-0.005 & 0.003 & 0.001 & 0.001 & 0.005 & 0.009 & 0.011\\
Ca & 0.105 & 0.060 &0.005  & -0.001& 0.004 & 0.000 & 0.007 & 0.006 & 0.009\\
Sc & 0.148 & 0.090 &0.007  & 0.000 & 0.000 &-0.001 & 0.007 & 0.004 & 0.008\\
Ti & 0.141 & 0.092 &0.008  & 0.000 & 0.003 & 0.000 & 0.008 & 0.003 & 0.009\\
V  & 0.110 & 0.054 &0.008  & 0.001 & 0.018 & 0.000 & 0.019 & 0.006 & 0.020\\
Cr & 0.098 & 0.057 &0.006  & -0.001& 0.004 & 0.000 & 0.007 & 0.004 & 0.008\\
Mn & 0.117 & 0.075 &0.008  & -0.001& -0.047& 0.000 & 0.048 & 0.007 & 0.048\\
Fe & 0.101 & 0.054 &0.006  & -0.001& 0.005 & 0.000 & 0.007 & 0.002 & 0.008\\
Co & 0.126 & 0.094 &0.008  & 0.001 & -0.064& 0.000 & 0.065 & 0.006 & 0.065\\
Ni & 0.120 & 0.072 &0.004  & 0.000 & 0.004 & 0.001 & 0.006 & 0.004 & 0.007\\
Cu & 0.129 & 0.088 &0.008  & 0.001 & -0.035& 0.001 & 0.035 & 0.008 & 0.036\\
Zn & 0.069 & 0.050 &0.001  & 0.001 & 0.006 & 0.002 & 0.006 & 0.043 & 0.043\\

\hline                                 
\end{tabular}
    \tablenotetext{1}{Errors due to stellar parameters}
    \tablenotetext{2}{Observational errors}
    \tablenotetext{3}{Quadric sum of the observational and stellar parameters uncertainties}
\end{minipage}
\end{table}

%% This figure uses \includegraphics to scale and rotate the still frame


\begin{thebibliography}{}

\bibitem[Bakos et al.(2007)]{bak07} Bakos, G.~{\'A}., Noyes, 
R.~W., Kov{\'a}cs, G., et al.\ 2007, \apj, 656, 552 


\bibitem[Chambers (2010)]{cham10} Chambers, J. E.\ 2010 \apjl, 724, 92 

\bibitem[Castelli \& Kurucz(2004)]{cas04} Castelli, F., \& Kurucz, R.~L.\ 2004, arXiv:astro-ph/0405087 

\bibitem[Cochram \& Hatzes(1997)]{coc97} Cochran, W., Hatzes, A., Butler, P,\& Marcy, G.\ 2009, \apjl, 483, 457 

\bibitem[Cutri et al.(2003)]{cut03} Cutri, R.~M., Skrutskie, 
M.~F., van Dyk, S., et al.\ 2003, yCat, 2246, 0 

\bibitem[Desidera et al.(2004)]{des04} Desidera, S., Gratton, R.G., Endl, M., et al. \ 2004, \aap, 420, 683

\bibitem[Desidera et al.(2006)]{des06} Desidera, S., Gratton, R.G., Lucatello, S., \& Claudi, R.U. \ 2006, \aap, 454, 581

%\bibitem[Fortney \& Nettelmann(2010)]{for10} Fortney, J.~J., \& Nettelmann, N.\ 2010, \ssr, 152, 423 

\bibitem[Gonzalez(1998)]{gon98} Gonzalez, G.\ 1998, \aap, 334, 221 

\bibitem[Gratton et al.(2001)]{gra01} Gratton R.G., Bonanno G., Claudi R.U., et al. \ 2001, \aap, 377, 123

\bibitem[Guillot(2005)]{gui05} Guillot, T.\ 2005, AREPS, 33, 493 

\bibitem[H{\o}g et 
al.(2000)]{hog00} H{\o}g, E., Fabricius, C., Makarov, V.~V., et al.\ 2000, \aap, 355, L27 

\bibitem[Laws \& Gonzalez(2001)]{la01} Laws, C., \& Gonzalez, G.\ 2001, \apj, 553, 405 

\bibitem[Liu et al.(2014)]{liu14} Liu, F., Asplund, M., 
Ram{\'{\i}}rez, I., Yong, D., \& Mel{\'e}ndez, J.\ 2014, \mnras, 442, L51 

\bibitem[Mack et al.(2014)]{mac14} Mack, C.~E., III, Schuler, 
S.~C., Stassun, K.~G., Pepper, J., \& Norris, J.\ 2014, arXiv:1404.1967

\bibitem[Mel{\'e}ndez et al.(2009)]{mel09} Mel{\'e}ndez, J., Asplund, M., Gustafsson, B., \& Yong, D.\ 2009, \apjl, 704, L66

\bibitem[Mel{\'e}ndez et 
al.(2012)]{mel12} Mel{\'e}ndez, J., Bergemann, M., Cohen, J.~G., et al.\ 2012, \aap, 543, A29 

\bibitem[Mermilliod et 
al.(1997)]{mer97} Mermilliod, J.-C., Mermilliod, M., \& Hauck, B.\ 1997, \aaps, 124, 349 


\bibitem[Monroe et al.(2013)]{mon13} Monroe, T.~R., 
Mel{\'e}ndez, J., Ram{\'{\i}}rez, I., et al.\ 2013, \apjl, 774, L32 

%\bibitem[Movshovitz et al.(2010)]{mov10} Movshovitz, N., Bodenheimer, P., Podolak, M., \& Lissauer, J.~J.\ 2010, \icarus, 209, 616 

\bibitem[Papaloizou 
\& Terquem(2006)]{pap06} Papaloizou, J.~C.~B., \& Terquem, C.\ 2006, RPPh, 69, 119 

\bibitem[Pl{\'a}valov{\'a} 
\& Solovaya(2013)]{pla13} Pl{\'a}valov{\'a}, E., \& Solovaya, N.~A.\ 2013, \aj, 146, 108 

\bibitem[Ram{\'{\i}}rez 
\& Mel{\'e}ndez(2005)]{ram05} Ram{\'{\i}}rez, I., \& Mel{\'e}ndez, J.\ 2005, \apj, 626, 465 

\bibitem[Ram{\'{\i}}rez et al.(2009)]{ram09} Ram{\'{\i}}rez, I., Mel{\'e}ndez, J., \& Asplund, M.\ 2009, \aap, 508, L17 

\bibitem[Ram{\'{\i}}rez et al.(2011)]{ram11} Ram{\'{\i}}rez, 
I., Mel{\'e}ndez, J., Cornejo, D., Roederer, I.~U., \& Fish, J.~R.\ 2011, \apj, 740, 76 

%\bibitem[Saumon \& Guillot(2004)]{sau04} Saumon, D., \& Guillot, T.\ 2004, \apj, 609, 1170 

\bibitem[Schuler et al.(2011)]{sch11} Schuler, S.~C., Cunha, 
K., Smith, V.~V., et al.\ 2011, \apjl, 737, L32 

\bibitem[Sneden(1973)]{sne73} Sneden, C.~A.\ 1973, Ph.D.~Thesis, 

\bibitem[Takeda(2005)]{tak05} Takeda, Y.\ 2005A, \pasj, 57, 83

\bibitem[Taylor(1986)]{tay86} Taylor, B.~J.\ 1986, \apjs, 60, 577 

\end{thebibliography}
\end{document}